# Metal Adsorption and Nucleation on Free-Standing Graphene by Low-Energy Electron Point Source Microscopy

*Marianna Lorenzo\*, Conrad Escher, Tatiana Latychevskaia, Hans-Werner Fink*

Physics Department, University of Zurich,

Winterthurerstrasse 190, 8057 Zurich, Switzerland

*Corresponding author: marianna@physik.uzh.ch

ABSTRACT

The interaction of metals with carbon materials, specifically with graphene, is of importance for various technological applications. In particular, intercalation of alkali metals is believed to provide a means for tuning the electronic properties of graphene for device applications. While the macroscopic effects of such intercalation events can readily be studied, following the related processes at an atomic scale in detail and under well-defined experimental conditions constitutes a challenge. Here, we investigate in situ the adsorption and nucleation of the alkali metals K, Cs, and Li on free-standing graphene by means of low-energy electron point source microscopy. We find that alkali metals readily intercalate in between bilayer graphene. In fact, the equilibrium distribution of K and Cs favours a much higher particle density in between the bilayer than on the single layer graphene. We obtain a quantitative value for the difference of the free energy of binding between these two domains. Our study is completed with a control experiment introducing Pd as a representative of the non-alkali metals. Now, we observe cluster formation in equal measure on both, single and bilayer graphene, however no intercalation. Our investigations thus constitute the first in situ study of metal atom sorption of different specificity on free-standing graphene.

TEXT

### Introduction

The properties of graphene,[1,2] i.e. an atomically thin carbon lattice, are expected to be valuable for a range of novel applications in various fields of technology. However, in order to be employed in electronic devices graphene needs functionalisation for tuning the band gap and hence adjusting the carrier concentration.[3,4] One approach thereto involves metal atom adsorption.[5] Of particular interest is doping graphene with alkali metals since they readily transfer their s-electron to graphene. The result can be pictured as a quasi-ionic bond between the alkali ion and the delocalized electron cloud in the graphene lattice. Upon formation of this ionic bond, the Fermi level shifts in the range from 0.88 to 1.29 eV.[4,6–8] Alkali metal doped graphene is therefore considered one of the most promising systems in electronics[9] and energy storage applications.[9–12] On the contrary, the adsorption of a transition metal atom on graphene results in the formation of a covalent bond with an electron density localized around the adatom.[13]

In general, type and strength of the bond between metal adsorbate and graphene together with the lateral interaction among adatoms drive the nucleation and growth mode. In the case of alkali metals, the dipole-dipole interaction between atoms leads to a two-dimensional growth, at least for sub-monolayer coverage,[7,13,14] while most of the transition metals tend to form three-dimensional clusters already at low coverage.[13,15,16]

For multiple layer systems, metal intercalation provides an additional route for tuning the electronic properties of carbon materials. In the case of graphite, intercalation is initiated at the layer edges, as illustrated in Figure 1a and proceeds by diffusion along the inter-planes.[17] Experiments have shown that except for Na all alkali metals readily intercalate after deposition on graphite (0001) at room temperature.[17,18] For supported graphene, experiments have shown that alkali metals intercalate between graphene and the substrate.[8,19–23] A dedicated experiment on Ir supported graphene has demonstrated that Cs and Li intercalate by migration through graphene vacancies or cracks localized at wrinkle crossings.[20] This intercalation pathway, schematically illustrated in Figure 1b, has been reported in several experimental studies.[23] On the contrary, when evaporated onto pristine graphene on SiC, intercalation is observed for Li only.[8,21,22] Theoretical studies suggest an intercalation pathway for Li via interim substitution of a carbon atom, as illustrated in Figure 1c. In fact, at high Li concentrations, the energy barrier for the formation of such temporary defect in the graphene lattice amounts to 0.8 eV. For all other alkali metals, this energy barrier is much higher (i.e., 4 eV for Cs), indicating that the penetration by substitution is an unlikely intercalation mechanism.[24] However, in view of theoretical predictions based on density functional theory (DFT) calculations, one should keep in mind that neither the dynamics of the process nor the entropy contribution to the free energy is included in such zero temperature models. Values for activation barriers accurate to a fraction of 1 eV might thus not always be expected from such models.

Here, we apply Low-Energy Electron Point Source (LEEPS) microscopy to investigate in situ adsorption and intercalation of alkali metals on free-standing graphene. For this purpose, openings are milled in a platinum metal coated SiN membrane and are covered with single layer graphene exhibiting few islands of bilayer graphene. The transition between these two different domains can be adequately described by a step edge. It is thus conceivable that alkali metal intercalation into the bilayer proceeds in a similar manner than in graphite, namely by diffusion through this single step edge, as depicted in Figure 1d. In the typical electron energy range of 30 to 250 eV graphene exhibits a transparency of about 70% per layer. Hence both, single as well as bilayer graphene can readily be inspected by LEEPS microscopy.[25,26] Moreover, low-energy electrons are extremely sensitive to localized charges and have been employed for imaging charged impurities.[27] Positive charges attract electrons whose trajectories are deflected towards the centre of the object and overlap resulting in a bright spot on the detector. Simulations have shown that the intensity of the spot increases with the amount of charge, and that fractions of an elementary charge can be detected.[27] The signature of both an individual charged adsorbate or a small charged cluster of adsorbates is a bright spot formed by the electrons deflected by the charge. This property provides means for detecting the positively charged alkali atom upon adsorption and subsequent electron transfer to graphene. Individual adsorption events, as well as a following nucleation of metals on free-standing graphene, can thus be observed in situ with a video frame rate time resolution and under ultra-high vacuum conditions. At low coverage, the repulsive dipole-dipole interaction prevents the formation of clusters, allowing to observe individual alkali metal ions. From the data obtained in this investigation, no difference in signature between adsorbed and intercalated alkali atoms in the bilayer domain could be identified. However, the analysis of the bright spot densities in the single layer and bilayer graphene domains provides indirect evidences for alkali metal intercalation. As a control experiment, next to alkali metal deposition, we also address the nucleation of Pd as a representative of the transition metals.

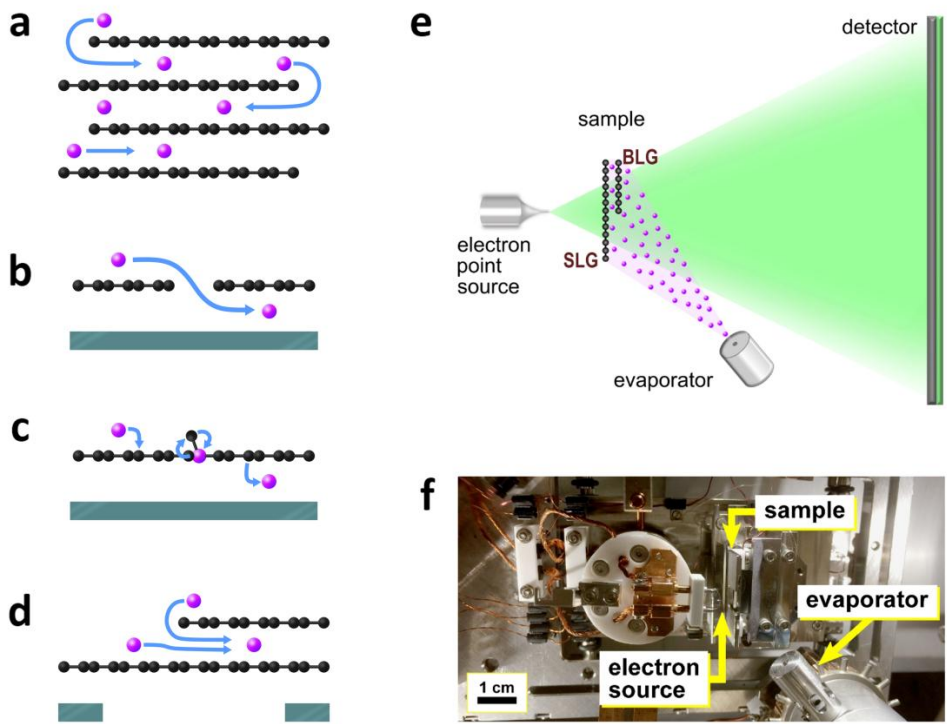

**Figure 1.** Illustration of the various intercalation pathways at left as well as a scheme and an image of the Low-Energy Electron Point Source microscope at right. (a) Intercalation through step edges in graphite (0001).[17,18] (b) Intercalation through a vacancy in supported graphene.[19,20,23] (c) Intercalation by carbon atom substitution.[23] Such process has been predicted only for Li.[24] (d) Intercalation through a step edge on free-standing graphene. (e) The electron point source is kept at a negative voltage ranging typically between 30 V and 250 V. Its relative

position to the sample can be adjusted in the nanometer range by means of piezo driven positioners. Typical source to sample distances range between 0.5 and 1 µm. The detector unit, a MCP detector followed by a phosphor screen, is placed 158 mm from the sample. The image formed on the screen is acquired by a CMOS sensor. Typical magnifications are of the order of $M \sim 10^5$. The evaporation of alkali metals is performed by means of commercial dispensers (SAES Getters) located 5 cm from the sample. (f) Top view of the optical bench.

**Experimental setup**

The LEEPS microscope[28] is designed for electron holography using the in-line scheme invented by Dennis Gabor.[29] Since it is a lens-less microscope, it relies on a coherent electron source based on an ultra-sharp W(111) tip, as illustrated in Figure 1e-f. Typical electron energies in the range of 30 to 250 eV correspond to wavelengths between 0.22 and 0.08 nm. The spherical electron wave originating from the point source scatters off the sample elastically (object wave) and interferes with the non-scattered wave (reference wave), producing an interference pattern (hologram) on the detector. Holograms are numerically reconstructed, in fact the amplitude and the phase distribution of the object wave can be retrieved.[30] The detector unit consists of a 75 mm diameter high-resolution multi-channel plate (MCP), followed by a phosphor screen and a CMOS sensor capable of acquiring holograms at a video rate of 25 fps. The distance $D$ between electron source and detector amounts to 158 mm and typical electron source to sample distances $d$ for imaging are between 0.5 and 1 µm, corresponding to a magnification of $M=D/d\sim 10^5$. Prior to adsorption studies, the system is baked out to achieve ultra-high vacuum conditions with a pressure in the low $10^{-9}$ mbar regime.

Alkali metals are evaporated in situ from commercial sources manufactured by SAES Getters.[31] These dispensers are activated through resistive heating with an evaporation rate increasing with temperature respectively heating current. Typical current values for operation are: 6.5 A for K, 7.6 A for Li, and 5 A for Cs. Once a stable evaporation rate is established, a mechanical shutter is opened, and the evaporation towards the sample continues for about 250 s. The deposition is performed at room temperature while maintaining a pressure in the low $10^{-8}$ mbar range during evaporation.

The samples are prepared using chemical vapor deposition (CVD) grown graphene on a copper substrate from ASC Material, Ltd. First, the graphene flake needs coating with poly(methyl methacrylate) (PMMA). The metal substrate can then be removed by wet etching in ammonium persulfate. After careful rinsing, the flake is transferred across an array of 500 nm wide square-cut openings in a Pd coated SiN membrane. Finally, the sample undergoes heat treatment so that the PMMA is catalytically removed as described elsewhere.[32]

In the course of the CVD growth of single layer graphene, there are to some extent islands of an additional layer forming on top of the continuous first layer. With this, monatomic step edges are present. In order to have access to these step edges the samples are inserted into the microscope with the islands facing the evaporator, as illustrated in Figure 1e.

**Results and discussion**

The signature of an alkali metal atom adsorbed on graphene turns out to be a bright spot in the hologram and thus provides evidence for the adsorbate being positively charged.[27] However, also for alkali metal intercalated or adsorbed within a bilayer island we observe the very same signature, be it for individual atoms or small clusters. Accordingly, this entails a certain loss of discriminatory power, but in turn makes counting of separate entities more reliable.

**K and Li deposition**

Alkali metal deposition on adjacent domains of single and bilayer graphene is monitored simultaneously either at a rate of 25 fps or by acquiring subsequent individual images with a delay time of approximately 5 s. The temporal evolution of the K coverage on single and bilayer graphene is presented in Figure 2a-b, and a movie of K deposition made of a sequence of 23 images acquired with a delay time of 5 s and an exposure time of 0.5 s is reported in the Supporting Information. A quantitative estimate of the particle density per unit area, $\rho_K$, as a function of the evaporation time is plotted in Figure 2b. The error on the number of particles N has been calculated as the sum of the statistical error ($\sqrt{N}$) and the error on the identification of the bright spots. As evident from this graph, the concentration of K within the bilayer domain is higher in comparison to that on the single layer. At the time $t = 87$ s, the density of particles within the bilayer is nearly three times higher than on the single layer.

Findings concerning related systems indicate a minimal coverage needed for intercalation via vacancies to take place.[20] This critical coverage corresponds to a density of adsorbates two orders of magnitude higher than in our experiment. Moreover, experiments on graphite report that intercalation through vacancy defects is possible only at high temperature, while at room temperature intercalation proceeds through step edges on the basal plane.[33] We assume that here the predominant intercalation pathway for K is via diffusion through the step edge, as illustrated in Figure 1d.

The temporal evolution of the Li coverage is shown in Figure 2c-d, together with the plot of the Li particle density $\rho_{Li}$ as a function of time. The progress is similar to that observed for K with a pronounced increase in the Li population in the bilayer domain over time for densities above $\rho_{Li} \approx 8\times10^{-4}$ $N_{Li}/nm^2$. At $t = 160$ s, the particle density within the bilayer is higher by a factor of about 2.3 compared to the single layer. These findings are in qualitative agreement with DFT calculations for the bilayer predicting a lower total energy for a Li atom between the layers in comparison to a Li atom adsorbed on the surface.[34] Concerning the pathway, it is conceivable that migration of Li through the step edge is the most likely one.

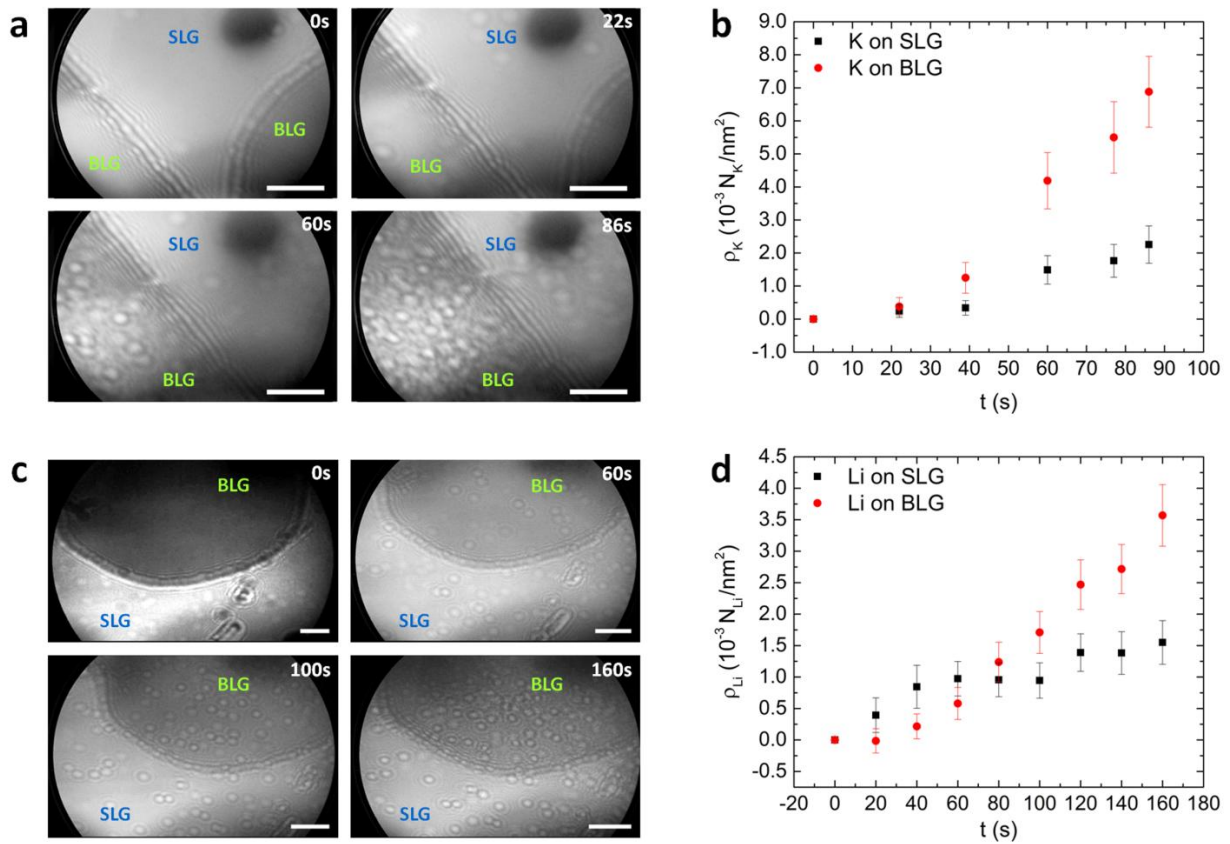

**Figure 2.** Temporal evolution of K and Li coverage on free-standing single layer graphene (SLG) and bilayer graphene (BLG). (a) Holograms of K adsorbates acquired at different times

with an electron energy in the range of 60 eV. The scale bar corresponds to 50 nm. (b) Particle density of K on SLG and on BLG as a function of time. (c) Holograms of Li adsorbates on SLG and BLG acquired at different times with an electron energy of 80 eV. The scale bar corresponds to 50 nm. (d) Particle density of Li on SLG and on BLG as a function of time.

A movie illustrating the build-up of the Li coverage with 25 fps is shown in the Supporting Information section. The movie shows apparent little diffusion for Li adatoms, while the particle density in the bilayer domain increases. Our view is that the mobility on the single layer should indeed be much higher than in between the bilayer. The visible atoms on the single layer are most likely localized at or around defects in graphene and carry out limited hopping around these "traps". The ones newly adsorbed form the vapour phase onto the single layer (as well as on top of the bilayer) are probably moving very rapidly and are not detectable with 25 fps before they reach the step to the bilayer domain and make it in between the two layers where, as a consequence, the density apparently increases. The same is probably true for those atoms that reach the domain boundary at a descending step from the top of the bilayer. Under these considerations, the interpretation of the temporal evolution data reported in Figure 2 is as follows: the number of detected particles in the single layer domain might be less than the number of adsorbed atoms. Assuming that the diffusion of the adsorbates is similar in both domains, $\rho_{SLG}$ coincides with the density of detected adsorbates in the bilayer domain. Given this, $\rho_{SLG}$ is the adsorbates' contribution to the total particle density $\rho_{BLG}$. The faster increase of $\rho_{BLG}$ with respect to $\rho_{SLG}$ during the deposition, reported in Figure 2, provides evidence that the excess of particles measured in the bilayer domain is due to intercalated ions.

During deposition, the evolution of the particle distribution appears similar for both, K and Li sorbates. However, once the deposition is stopped, the two systems evolve significantly different. In Figure 3, two pairs of holograms are shown; one concerns the deposition of K (Figure 3a- b) the other one concerns the deposition of Li (Figure 3c-d). The first hologram of each pair is acquired about 200 s after having started the deposition. The second hologram represents the final distribution after the deposition has been stopped. From the comparison of the two graphene domains in Figure 3a and Figure 3b it is evident that once the deposition is terminated, the K coverage of the single layer is clearly thinning out. The result is an unequal degree of coverage in the two domains. In fact, the particle densities amount to $\rho_{SLG} = 3.9 \times 10^{-4}$ $N_K/nm^2$ and $\rho_{BLG} = 9 \times 10^{-3}$ $N_K/nm^2$ for the single and the bilayer graphene, respectively. Under the assumption that thermodynamic equilibrium has been achieved and that transitions between the two states occur while not changing the average densities, the ratio between $\rho_{SLG}$ and $\rho_{BLG}$ is related to the difference in the free energy of binding $\Delta E$ between the two binding states, on the single layer respectively in the bilayer, through the Boltzmann factor:

$$\frac{\rho_{SLG}}{\rho_{BLG}} = e^{\frac{-\Delta E}{k_B T}}$$

whereby $k_B$ is the Boltzmann constant and $T$ the absolute temperature. From this relation, we calculate a value of $\Delta E = 0.08 \pm 0.01$ eV, which indicates that the adsorption on the single layer is energetically less favourable than intercalation within the bilayer domain. This value should be considered as a lower limit for the free energy difference of binding since the analysis assumes that there are just two different binding states for the K atom to adopt, either adsorbed on the graphene surface or intercalated in between two layers. However, there might be some trapping sites for K atoms on the single layer domain, representing a third binding state. In this case, our analysis would lead to an under-estimation of the free energy difference between the adsorption on ideal single layer graphene and the intercalation in between two layers, since we assumed no competing alternative sites, represented by some likely defects on the otherwise almost perfect and clean graphene layer.

In contrast to the observations described above, in the case of Li the coverage is preserved on both graphene domains after terminating the deposition, as evident from Figure 3c-d. Moreover, for the adsorbates on the single layer diffusion could not be observed. Instead, localized clusters of different shapes and sizes seem to have rapidly formed. Consequently, no sensible estimate for the difference in the free energy of binding can be carried out for these data acquired at room temperature. Cluster formation, however, is in line with DFT simulations predicting that the repulsive dipole-dipole interaction may be overcome at high coverage. With increased coverage, the activation energy for cluster formation is thus reduced leading to the formation of stable Li clusters.[35–37]

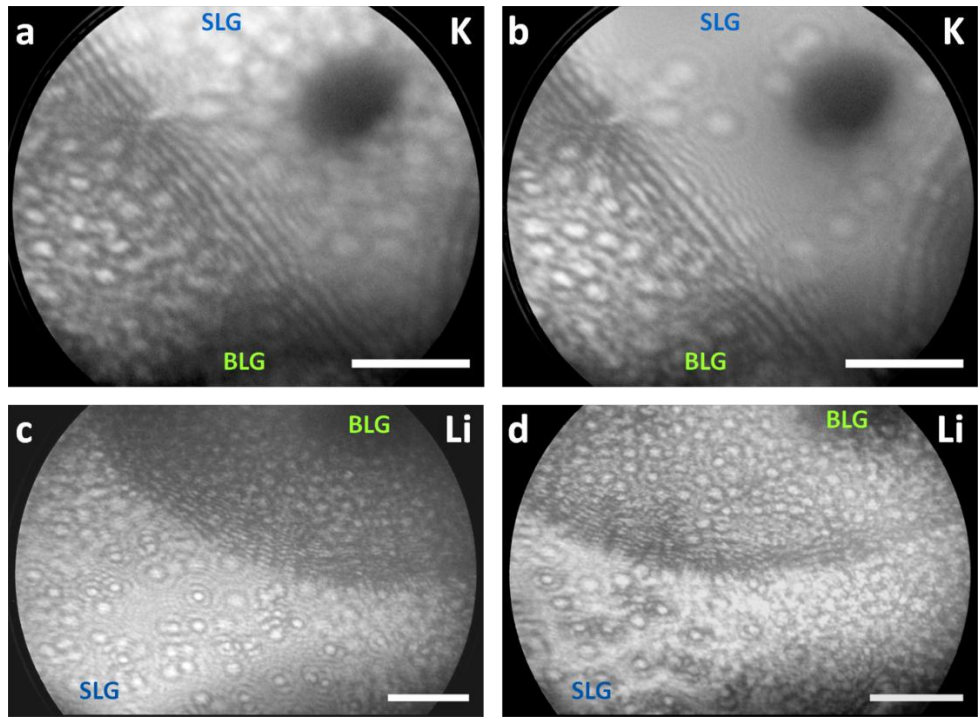

**Figure 3.** Distribution of K and Li adsorbates on free-standing single layer (SLG) and bilayer (BLG) graphene after deposition. (a) Distribution of K adsorbates after continuous deposition for 136 s imaged with 54 eV electrons and (b) 90 s after the termination of deposition imaged with 57 eV electrons. (c) Distribution of Li adsorbates after continuous deposition for 220 s and (d) 60 s after the termination of deposition. Both holograms are acquired at an electron energy of 80 eV. The scale bar corresponds to 50 nm.

### Cs deposition

For the deposition of Cs similar characteristics are observed compared to the ones reported on K. Again, the final distribution after the deposition has been stopped presents a higher particle density $\rho_{Cs}$ in bilayer graphene than on the single layer, as apparent in Figure 4b. With $\rho_{Cs} = 3.3 \times 10^{-2}$ $N_{Cs}/nm^2$ in the bilayer the density is comparable to the corresponding value found for

K. The ratio between the degrees of coverage within the bilayer and the single layer domain respectively, and with that the difference in the free energy of binding between the two domains, might even be larger in the case of Cs because the only Cs adsorbates found on the single layer have anchored to residual adsorbates stemming from sample preparation. Subsequently, Cs nucleation proceeds around these pre-existing adsorbates and forms positively charged clusters. The latter is indicated by the pronounced bright spots at the nucleation sites in the single layer domain of Figure 4b.

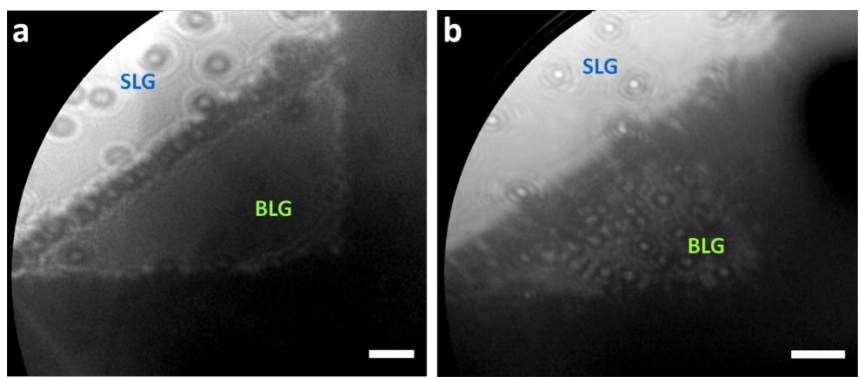

**Figure 4.** Cs adsorption on free-standing single layer graphene (SLG) and bilayer graphene (BLG). (a) Graphene sample before Cs deposition showing pre-existing adsorbates, located mainly on the SLG, and (b) 90 s after Cs deposition for 5 min. Apparently, Cs atoms cluster around these pre-existing adsorbates on the SLG. In the BLG, a fairly high density of intercalated Cs is observed. Both holograms are acquired with 100 eV electrons. The scale bar corresponds to 50 nm.

**Pd deposition**

In a control experiment, Pd is introduced as a representative of the non-alkali transition metals. Now, cluster formation occurs in equal measure on the single and the bilayer domain during deposition on free-standing graphene. Low energy electron holograms of Pd deposited on single and bilayer graphene are shown in Figure 5b. In contrast to the alkali metal sorbates, the Pd adsorbates do not appear as bright spots in the holographic record. This implies that the adsorbates are not charged and regular reconstruction routines as described elsewhere are applicable.[30] The corresponding amplitude and phase reconstructions are displayed in Figure 5c. Based on these reconstructions, we measure cluster diameters ranging from 22 to 60 nm.

The formation of such clusters agrees with previously reported experimental results. Scanning probe measurements have proven that Pd evaporated on highly-oriented pyrolytic graphite (HOPG)[38] as well as on supported graphene[15,16] at room temperature leads to the formation of spheroidal clusters. Howsoever, our experiment does not provide information about the three-dimensionality of the Pd clusters formed on free-standing graphene, nor does it contain evidence for intercalation of Pd in the bilayer.

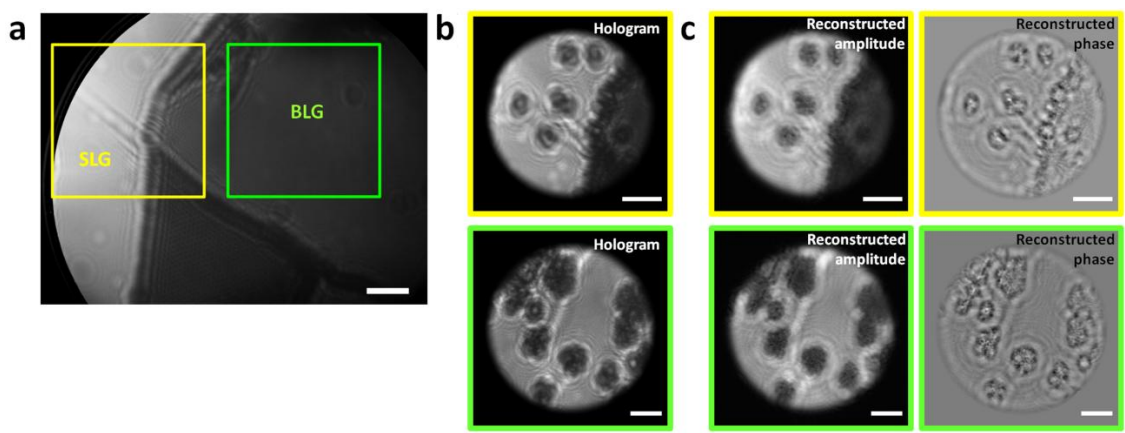

**Figure 5.** Pd adsorption and nucleation on free-standing single layer graphene (SLG) and bilayer graphene (BLG). (a) Graphene window prior to deposition with distinct areas of SLG and BLG.

The hologram is acquired at an electron energy of 80 eV. (b) Holograms of Pd clusters on the SLG, corresponding to the area indicated by the yellow square in (a), and on the BLG, corresponding to the area indicated by the green square in (a). Both holograms are acquired at an electron energy of 70 eV. (c) Amplitude and phase reconstructions of the holograms shown in (b), reveal cluster formation. The scale bar corresponds to 50 nm.

**Conclusion and Outlook**

By employing a dedicated LEEPS microscope and taking advantage of the sensitivity of low-energy electrons to localized charges, it became possible to in situ study adsorption, nucleation and intercalation of alkali metals on free-standing graphene. In the case of K, quantitative values of the energetics involved could be obtained at room temperature. In view of further investigations aiming at the energetics associated with intercalation versus adsorption for all alkali metals on an atomic scale, extending the temperature range seems desirable. Moreover, it seems conceivable that local microscopic information could in the future be combined with electron transport data. By electrically contacting the graphene sample for a four-point measurement of its conductivity, it should be possible to correlate conduction with image information on adsorbed or intercalated atoms. The scattering of conduction electrons at atomic sites might even manifest itself in the associated images due to a potential drop across graphene.

ASSOCIATED CONTENT

**Supporting Information**.

The following files are available free of charge at http://pubs.acs.org.

- Movie clip of Li deposition on adjacent domains of single and bilayer graphene acquired at an electron energy of 80 eV. The scale bar corresponds to 50 nm. (AVI)

- Movie clip of K deposition on adjacent domains of single layer and bilayer graphene acquired at an electron energy of 60 eV. The movie has been assembled from a sequence of images acquired with a delay time of 5 s, and it has speeded up to 10X. The scale bar corresponds to 50 nm. (AVI)

- Comparison between LEEPS microscopy and TEM images. (PDF)

AUTHOR INFORMATION

**Corresponding Author**

* Marianna Lorenzo, marianna@physik.uzh.ch

**Author Contributions**

M. L. performed the experiments and the data analysis. C.E. and H.-W. F. developed the low-energy electron point source microscope used in this study. T.L. performed the hologram reconstructions and the data analysis check. The manuscript was written through contributions of all authors. All authors have given approval to the final version of the manuscript.

**Funding Sources**

Swiss National Science Foundation, grant number 200021- 163453.

ACKNOWLEDGMENT

This study has been supported by the Swiss National Science Foundation (SNF Grant No. 200021- 163453).

CONFLICT OF INTEREST

The authors declare no conflict of interest.

ABBREVIATIONS

BLG, bilayer graphene; CVD, chemical vapour deposition; DFT, density functional theory; LEEPS, low-energy electron point source; MCP, multi-channel plate; PMMA, Poly(methyl methacrylate); SLG, single layer graphene.

REFERENCES

(1)    Novoselov, K. S.; Geim, A. K.; Morozov, S. V; Jiang, D.; Zhang, Y.; Dubonos, S. V; Grigorieva, I. V; Firsov, A. A., "Electric Field Effect in Atomically Thin Carbon Films.", *Science* **2004**, *306* (5696), 666–669. DOI: 10.1126/science.1102896

(2)    Geim, A. K.; Novoselov, K. S., "The Rise of Graphene.", *Nat. Mater.* **2007**, *6* (3), 183–191. DOI: 10.1038/nmat1849

(3)    Liu, H.; Liu, Y.; Zhu, D., "Chemical Doping of Graphene.", *J. Mater. Chem.* **2011**, *21* (10), 3335–3345. DOI: 10.1039/C0JM02922J

(4)    Haberer, D.; Petaccia, L.; Fedorov, A. V.; Praveen, C. S.; Fabris, S.; Piccinin, S.; Vilkov, O.; Vyalikh, D. V.; Preobrajenski, A.; Verbitskiy, N. I.; Shiozawa, H.; Fink, J.; Knupfer, M.; Büchner, B.; Grüneis, A., "Anisotropic Eliashberg Function and Electron-Phonon Coupling in Doped Graphene.", *Phys. Rev. B* **2013**, *88* (8), 81401. DOI:

10.1103/PhysRevB.88.081401

(5) Uchoa, B.; Lin, C.-Y.; Castro Neto, A. H., "Tailoring Graphene with Metals on Top.", *Phys. Rev. B* **2008**, *77* (3), 35420. DOI: 10.1103/PhysRevB.77.035420

(6) Bianchi, M.; Rienks, E. D. L.; Lizzit, S.; Baraldi, A.; Balog, R.; Hornekær, L.; Hofmann, P., "Electron-Phonon Coupling in Potassium-Doped Graphene: Angle-Resolved Photoemission Spectroscopy.", *Phys. Rev. B* **2010**, *81* (4), 41403. DOI: 10.1103/PhysRevB.81.041403

(7) Chan, K. T.; Neaton, J. B.; Cohen, M. L., "First-Principles Study of Metal Adatom Adsorption on Graphene.", *Phys. Rev. B* **2008**, *77* (23), 1–12. DOI: 10.1103/PhysRevB.77.235430

(8) Watcharinyanon, S.; Virojanadara, C.; Johansson, L. I., "Rb and Cs Deposition on Epitaxial Graphene Grown on 6H-SiC(0001).", *Surf. Sci.* **2011**, *605* (21–22), 1918–1922. DOI: 10.1016/j.susc.2011.07.007

(9) Rezapour, M. R.; Myung, C. W.; Yun, J.; Ghassami, A.; Li, N.; Yu, S. U.; Hajibabaei, A.; Park, Y.; Kim, K. S., "Graphene and Graphene Analogs toward Optical, Electronic, Spintronic, Green-Chemical, Energy-Material, Sensing, and Medical Applications.", *ACS Appl. Mater. Interfaces* **2017**, *9* (29), 24393–24406. DOI: 10.1021/acsami.7b02864

(10) Buldum, A.; Tetiker, G. J., "First-Principles Study of Graphene-Lithium Structures for Battery Applications.", *Appl. Phys.* **2013**, *113* (15), 154312. DOI: 10.1063/1.4802448

(11) Fan, X.; Zheng, W. T.; Kuo, J.-L.; Singh, D. J., "Adsorption of Single Li and the Formation of Small Li Clusters on Graphene for the Anode of Lithium-Ion Batteries.",

*ACS Appl. Mater. Interfaces* **2013**, *5* (16), 7793–7797. DOI: 10.1021/am401548c

(12) Eftekhari, A.; Jian, Z.; Ji, X., "Potassium Secondary Batteries.", *ACS Appl. Mater. Interfaces* **2017**, *9* (5), 4404–4419. DOI: 10.1021/acsami.6b07989

(13) Liu, X.; Wang, C.-Z.; Hupalo, M.; Lin, H.-Q.; Ho, K.-M.; Tringides, M., "Metals on Graphene: Interactions, Growth Morphology, and Thermal Stability.", *Crystals* **2013**, *3* (1), 79–111. DOI: 10.3390/cryst3010079

(14) Jin, K.-H.; Choi, S.-M.; Jhi, S.-H., "Crossover in the Adsorption Properties of Alkali Metals on Graphene.", *Phys. Rev. B* **2010**, *82*, 4–7. DOI: 10.1103/PhysRevB.82.033414

(15) Yagyu, K.; Takahashi, K.; Tochihara, H.; Tomokage, H.; Suzuki, T., "Neutralization of an Epitaxial Graphene Grown on a SiC(0001) by Means of Palladium Intercalation.", *Appl. Phys. Lett.* **2017**, *110* (13), 131602. DOI: 10.1063/1.4979083

(16) Zhou, Z.; Gao, F.; Goodman, D. W., "Deposition of Metal Clusters on Single-Layer Graphene/Ru(0001): Factors that Govern Cluster Growth.", *Surf. Sci.* **2010**, *604* (13–14), L31–L38. DOI: 10.1016/j.susc.2010.03.008

(17) Dresselhaus, M. S.; Dresselhaus, G., "Intercalation Compounds of Graphite.", *Adv. Phys.* **2002**, *51* (1), 1–186. DOI: 10.1080/00018730110113644

(18) Caragiu, M.; Finberg, S., "Alkali Metal Adsorption on Graphite: a Review.", *J. Phys. Condens. Matter* **2005**, *17*, R995–R1024. DOI: 10.1088/0953-8984/17/35/R02

(19) O'Hern, S. C.; Stewart, C. A.; Boutilier, M. S. H.; Idrobo, J.-C.; Bhaviripudi, S.; Das, S. K.; Kong, J.; Laoui, T.; Atieh, M.; Karnik, R., "Selective Molecular Transport through Intrinsic Defects in a Single Layer of CVD Graphene.", *ACS Nano* **2012**, *6* (11), 10130–

10138. DOI: 10.1021/nn303869m

(20) Petrović, M.; Šrut Rakić, I.; Runte, S.; Busse, C.; Sadowski, J. T.; Lazić, P.; Pletikosić, I.; Pan, Z.-H. Z.-H.; Milun, M.; Pervan, P.; Atodiresei, N.; Brako, R.; Šokčević, D.; Valla, T.; Michely, T.; Kralj, M., "The Mechanism of Caesium Intercalation of Graphene.", *Nat. Commun.* **2013**, *4*, 2772. DOI: 10.1038/ncomms3772

(21) Virojanadara, C.; Zakharov, A. A.; Watcharinyanon, S.; Yakimova, R.; Johansson, L. I., "A Low-Energy Electron Microscopy and X-ray Photo-Emission Electron Microscopy Study of Li Intercalated into Graphene on SiC(0001).", *New J. Phys.* **2010**, *12* (12), 125015. DOI: 10.1088/1367-2630/12/12/125015

(22) Virojanadara, C.; Watcharinyanon, S.; Zakharov, A. A.; Johansson, L. I., "Epitaxial Graphene on 6H-SiC and Li Intercalation.", *Phys. Rev. B* **2010**, *82* (20), 205402. DOI: 10.1103/PhysRevB.82.205402

(23) Liu, X.; Han, Y.; Evans, J. W.; Engstfeld, A. K.; Behm, R. J.; Tringides, M. C.; Hupalo, M.; Lin, H.-Q.; Huang, L.; Ho, K.-M.; Appy, D.; Thiel, P. A.; Wang, C.-Z., "Growth Morphology and Properties of Metals on Graphene.", *Prog. Surf. Sci.* **2015**, *90* (4), 397–443. DOI: 10.1016/j.progsurf.2015.07.001

(24) Boukhvalov, D. W.; Virojanadara, C., "Penetration of Alkali Atoms throughout a Graphene Membrane: Theoretical Modeling.", *Nanoscale* **2012**, *4* (5), 1749. DOI: 10.1039/c2nr11892k

(25) Mutus, J. Y.; Livadaru, L.; Robinson, J. T.; Urban, R.; Salomons, M. H.; Cloutier, M.; Wolkow, R. a., "Low-Energy Electron Point Projection Microscopy of Suspended

Graphene, the Ultimate 'Microscope Slide'", *New J. Phys.* **2011**, *13*, 63011. DOI: 10.1088/1367-2630/13/6/063011

(26) Longchamp, J. N.; Latychevskaia, T.; Escher, C.; Fink, H. W., "Non-Destructive Imaging of an Individual Protein.", *Appl. Phys. Lett.* **2012**, *101*, 1–5. DOI: 10.1063/1.4752717

(27) Latychevskaia, T.; Wicki, F.; Longchamp, J. N.; Escher, C.; Fink, H. W., "Direct Observation of Individual Charges and Their Dynamics on Graphene by Low-Energy Electron Holography.", *Nano Lett.* **2016**, *16* (9), 5469–5474. DOI: 10.1021/acs.nanolett.6b01881

(28) Fink, H.-W.; Stocker, W.; Schmid, H., "Holography with Low-Energy Electrons.", *Phys. Rev. Lett.* **1990**, *65* (10), 1204–1206. DOI: 10.1103/PhysRevLett.65.1204

(29) Gabor, D., "A New Microscopic Principle.", *Nature* **1948**, *161* (4098), 777–778. DOI: 10.1038/161777a0

(30) Latychevskaia, T.; Fink, H.-W., "Practical Algorithms for Simulation and Reconstruction of Digital In-Line Holograms.", *Appl. Opt.* **2015**, *54* (9), 2424–2434. DOI: 10.1364/AO.54.002424

(31) SAES Getters, S.P.A, Italy. https://www.saesgetters.com/.

(32) Longchamp, J.-N.; Escher, C.; Fink, H.-W., "Ultraclean Freestanding Graphene by Platinum-Metal Catalysis.", *J. Vac. Sci. Technol. B* **2013**, *31*, 20605. DOI: 10.1116/1.4793746

(33) Büttner, M.; Choudhury, P.; Karl Johnson, J.; Yates, J. T., "Vacancy Clusters as Entry Ports for Cesium Intercalation in Graphite.", *Carbon N. Y.* **2011**, *49*, 3937–3952. DOI:

10.1016/j.carbon.2011.05.032

(34) O'Hara, A.; Kahn, R. E.; Zhang, Y. Y.; Pantelides, S. T., "Defect-Mediated Leakage in Lithium Intercalated Bilayer Graphene.", *AIP Adv.* **2017**, *7* (4), 45205. DOI: 10.1063/1.4980052

(35) Lee, E.; Persson, K. A., "Li Absorption and Intercalation in Single Layer Graphene and Few Layer Graphene by First Principles.", *Nano Lett.* **2012**, *12* (9), 4624–4628. DOI: 10.1021/nl3019164

(36) Liu, Y.; Artyukhov, V. I.; Liu, M.; Harutyunyan, A. R.; Yakobson, B. I., "Feasibility of Lithium Storage on Graphene and Its Derivatives.", *J. Phys. Chem. Lett.* **2013**, *4* (10), 1737–1742. DOI: 10.1021/jz400491b

(37) Liu, M.; Kutana, A.; Liu, Y.; Yakobson, B. I., "First-Principles Studies of Li Nucleation on Graphene.", *J. Phys. Chem. Lett.* **2014**, *5* (7), 1225–1229. DOI: 10.1021/jz500199d

(38) Nie, H. Y.; Shimizu, T.; Tokumoto, H., "Atomic Force Microscopy Study of Pd Clusters on Graphite and Mica.", *J. Vac. Sci. Technol. B* **1994**, *12* (3), 1843. DOI: 10.1116/1.587652